\documentclass[aps,twocolumn,showpacs,prb]{revtex4}

\usepackage{epsfig}

\newcommand{\bn}[1]{\mbox{\boldmath $#1$}}

\newcommand{\be}{\begin{equation}}
\newcommand{\ee}{\end{equation}}
\newcommand{\bc}{\begin{center}}
\newcommand{\ec}{\end{center}}
\newcommand{\bea}{\begin{eqnarray}}
\newcommand{\eea}{\end{eqnarray}}
\newcommand{\ba}{\begin{array}}
\newcommand{\ea}{\end{array}}
\newcommand{\nn}{\nonumber}

\begin{document}

\title{Optical transition in self-assembled InAs/GaAs quantum lens under high hydrostatic pressure}

\author{Arezky H. Rodr\'{\i}guez$^1$, C. Trallero-Giner$^2$, C. A. Duque$^3$ and G. J. V\'azquez$^4$}

\affiliation{$^{1}$ Academia de Matem\'aticas, Universidad Aut\'onoma de la Ciudad de M\'exico,
Mexico City, Mexico \\
$^2$ Facultad de F\'{\i}sica, Universidad de La Habana, 10400 C. Habana, Cuba. \\
$^3$ Instituto de F\'{\i}sica, Universidad de Antioquia, AA 1226, Medell\'{\i}n, Colombia. \\
$^4$ Instituto de F\'{\i}sica, Universidad Nacional Aut\'onoma de M\'exico (UNAM), Apdo. Postal
20-364, San \'Angel 01000, M\'exico D.F., Mexico.}

\date{\today}

\begin{abstract}
We present a simulation to characterize the dependence on hydrostatic pressure for the
photoluminescence spectra in self-assembled quantum dots with lens shape geometry. We have tested
the physical effects of the band offset and electron-hole effective masses on the optical emission
in dot lens. The model could be implemented to get qualitative information of the parameters
involved in the quantum dot or the measured optical properties as function of pressure.
\end{abstract}

\pacs{02.30.Nw,73.22.-f,73.22.Dj,81.40.Vw}

\keywords{}

\maketitle

\section{Introduction}
\label{sec:intro}

The study of the physical properties of nanostructures is of great importance nowadays.
Specifically, self-assembled quantum dots (QDs) grown by Stranski-Krastanov method
\cite{Stranski-Krastanow} have attracted great attention from the scientific community
\cite{hawrylak-book,Grundmann-book}. A lot of works have been devoted to the study of the
electronic and optical properties of QDs where their shape are modeled as quantum pyramids
\cite{cusack96,cusack97,grundmann95,zunger99,pryor98,grundmann99,zunger01,mervyn}. It has been also
considered QDs having lens shape
\cite{hawrylak96-1,eaglesham90,forchel96,hapke99,zhu98,liao98,moison94,arezky01,zou99,even}. In
this case, there is a rather extended opinion which states that the work done considering QDs with
pyramids shapes can be straightforward applied to characterize the QDs with lens shape and
therefore, the calculation made in pyramids QDs does not need to be re-calculated in quantum
lenses. Nevertheless, different authors have shown the importance to consider the full lens
geometry for a better  description  of the optical properties of the QDs
\cite{ngo,lew,AHR2006pss1,even}.

Photoluminescence (PL) measurement under high hydrostatic pressure has been proveed to be an
effective tool for exploring the electronic structure and optical transitions in QDs
\cite{tang,goni1,goni2,wei,Itskevich97,Itskevich98,Itskevich99,ma2004}. It is reported a
significant decreasing in the hydrostatic pressure coefficient (PC) for InAs/GaAs QDs in comparison
with bulk material. Different experimental works have proposed some explanations: non-uniform
Indium distribution in the QDs \cite{mintairov}, pressure dependence of the effective masses and
confined potential \cite{ma2004}, and the necessity to include a nonlinear strain distribution
\cite{goni1} and band-gap pressure dependence and finite barriers \cite{luo}. From the theoretical
point of view some works have been devoted to explain the obtained low value of hydrostatic PC. For
example, elasticity theory in Ref. \onlinecite{tang}, atomistic empirical pseudo-potential method
in Ref. \onlinecite{wei} and atomistic valence-force field method in Ref. \onlinecite{goni2} were
implemented. In all of the former works, the QDs have been modeled as quantum pyramids.

The motivation of the presente work is focused on the experimental observations of Ref.
\onlinecite{goni1,Itskevich97,Itskevich98,Itskevich99}. They have performed studies of
low-temperature PL in InAs and InGaAs self-\-assembled quantum dots (SAQDs) embedded in GaAs matrix
under hydrostatic pressure. From the pressure dependence of the main electron-hole transition
energy, a pressure coefficient 20\% smaller that the InAs case was reported. In Refs.
\onlinecite{tang} and \onlinecite{Itskevich97} some arguments could explain this difference in the
pressure coefficients: (i) the change of the band offset between InAs and GaAs, (ii) the strong
tension in the InAs SAQD material and, (iii) the lowering of the electron and light-hole
quantization energies due to the change of the effective masses with pressure.

In Ref. \onlinecite{Itskevich99} was argued that the dots can be characterized by a lens symmetry
with a base between $7.5$ and $10.6$ nm of radius and ratio height/radius between $0.28$ and $0.4$.
We need to appoint that, in general, the shape and dimensions of the InAs/GaAs SAQDs are not always
well determined and a degree of freedom remains.

An attempt to describe the PL peak energies can be found in Ref. \onlinecite{duque-jpcm} where the
spatial confined geometry was modelled as a parabolic cylinder potential avoiding the specific
geometry of the SAQDs. It is well known that different geometries yield different results, as shown
in Ref. \onlinecite{AHR2006pss1} when comparing the electronic properties between a cylindrical and
a lens shape SAQD. In this paper, the QD will be modeled by a full lens symmetry with maximum
height $b$ and circular cross section of radius $a$ with $b<a$. It is implemented a model which
allows the inclusion of the finite barrier effect in SAQD with lens shape. The model allows to
validate the importance of the change of the band offset and electron-hole effective masses as a
function of pressure on the PL spectra. In our simulation the strain effect within the QD has been
taking into account through an effective linear pressure coefficient at zero temperature which
consider the built-in strain of the dot by assuming a simple two-dimensional strained layers model
(see Table \ref{tabla1} and Refs. \onlinecite{ma2004,duque-jpcm,frogley}). The QDs considered here
are within the strong confinement regime and as first approximation the excitonic effects are
ignored.

The remainder of the paper is organized as follows. Section \ref{basic-relations} deals with the
general trends of the simulation for including the hydrostatic pressure in QDs characterized by a
lens geometry with a short discussion of the model used to obtain the electronic spectrum. Section
\ref{effect-of-the-pressure} presents theoretical calculations for the system comparing the effect
of the hydrostatic pressure over different quantum lenses. A comparison with available PL data of
the main transition energy as a function of $P$ is presented in Sec. \ref{section-fitting}.
Finally, section \ref{conclusions} is devoted to the conclusions.

\section{Basic relations}
\label{basic-relations}

The fractional change in volume in the InAs domain is given by
\be
\label{change_of_volume}
\delta V / V = - 3 P (S_{11} + 2 S_{12}),
\ee
where $S_{11} = 1.946 \times 10^{-3} \, kbar^{-1}$ and $S_{12} = - 6.855 \times 10^{-4} \,
kbar^{-1}$ are the compliance constants \cite{duque-jpcm}. The energy gap $E_g$ at low temperature
at the $\Gamma$ point is proportional to the pressure according to
\be
\label{Egap}
E_g(P,T) = E_g^{(o)} + \alpha \, P - \frac{\kappa \, T^2}{T + c},
\ee
where $E_g^{(o)}$ is the gap energy at $T=0$ and $P=0$, $\alpha$ is the pressure coefficient and
$\kappa$ and $c$ are the temperature coefficients \cite{duque-jpcm}. For the case of InAs/GaAs
SAQD, the parameters used for both materials are shown in Table \ref{tabla1}. With the inclusion of
the finite stress it can be seen that the $\alpha$ coefficient is greater for the GaAs material
than the InAs. Hence, the confined potential for the conduction and valence bands will increase
with pressure. The variation of the effective masses was evaluated in terms of the fundamental gaps
according to the Kane model and it was obtained \cite{trallero-unpublished}
\be
\label{mass_electron} \frac{m_o}{m^*_e} = 1 + 2 \left( \frac{P_o^2}{m_o} \right) \frac{E_g +
\frac{2}{3} \delta}{E_g \left( E_g + \delta \right)}, \ee for electrons, \be \label{mass_ll}
\frac{m_o}{m^*_{lh}} = - \left[ 1 - \frac{4}{3 \; E_g} \left( \frac{P_o^2}{m_o} \right) \right],
\ee
for light holes and
\be
\label{mass_hl}
\frac{m_o}{m^*_{hh}} = \gamma_1 - 2 \; \gamma_2
\ee
for the heavy holes. Here $E_g$ is the gap energy given by Eq. (\ref{Egap}), $\gamma_1$ and
$\gamma_2$ are the Luttinger parameters, $\delta$ is the spin-orbit splitting and $m_o$ is the free
electron mass. The parameter $P_o$ is the interband momentum matrix element between the conduction
and valence bands.

The PL lines in such QDs structures are proportional to the optical transition energies
$E_{N_e,m_e;N_h,m_h}(P)$ which are function of the pressure. The energy levels at a given band are
characterized by the radial label $N$ with axial projection of the angular momentum with quantum
number $m$. Then, in the framework of the parabolic band model, the transition energy
$E_{N_e,m_e;N_h,m_h}$ means the energy difference between the states ($N_h,m_h$) and ($N_e,m_e$) in
the valence and conduction bands, respectively. The Oscillator Strength (OS) is given by
\cite{PRB-DR}:
\be
\label{OS}
d(N_e, m_e;N_h,m_h) = \left| \int_D \Psi_{N_e,m_e;N_h,m_h} (\bn{r},\bn{r}) \mbox{d} \bn{r}
\right|^2.
\ee
In the case of a SAQD with axial symmetry, it is obtained for the former expression
$d(N_e,m_e;N_h,m_h) = 0$ for $m_e \ne m_h$ and only the terms with $m_e = m_h$ remains. The
integration is over the whole domain $D$ of the QD.

\subsection{Electronic structure}
\label{electronic_structure}

\begin{figure} 
\vspace*{0.5cm}
\centerline{\psfig{file=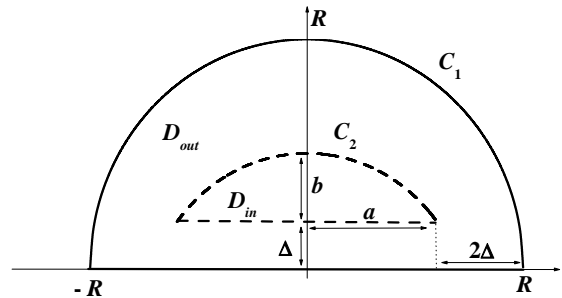,width=80mm}}
\vspace*{-2.5cm}
\caption{Transversal section of a 3D lens $D_{in}$ in a semispherical medium $D_{out}$ with
contours $C_2$ and $C_1$ respectively. Contour $C_1$ and $C_2$ are se\-pa\-ra\-ted a distance equal
or greater than $\Delta$ along the perpendicular axis.}
\label{fig:FinBarr}
\end{figure}

Although previous theoretical studies in QDs with lens shape \cite{jpacm,jap,PRB-DR} considered
infinite  potential wall, the finite value of the band offset is a fundamental parameter when
considering the hydrostatic pressure, as can be seen from Eq. (\ref{Egap}). Then, in our case, the
problem for a finite barrier will be modelled including a lens-shape well potential with height
$V_o$ in a hard-walls semi-spherical region, as shown in Fig. \ref{fig:FinBarr}. The semi-spherical
region is divided in two regions, $D_{in}$ and $D_{out}$ with confined potential zero and $V_o$,
respectively, and the mismatch boundary conditions are considered at the interface. Assuming
parabolic band dispersion for electron and holes, the solution of the effective mass equation for a
lens geometry, with finite barrier, in an infinite surrounding medium can be obtained by minimizing
the effect of the external boundary $C_1$ over the wavefunction of the corresponding energy level
under study. This can be achieved taken larger enough value of the distance $\Delta$. The equation
for the whole region $D$ is:
\bea
\label{eq1}
- \frac{\hbar^2}{2} \nabla \left( \frac{1}{m^*(r,\theta,\phi)} \nabla \Psi \right) +
V(r,\theta,\phi) \, \Psi = E \, \Psi,  \\
(r,\theta,\phi) \in D = D_{in} + D_{out}, \nn
\eea
where
$$
V(r,\theta,\phi) = \left\{
\begin{tabular}{ll}
0 & ; $(r,\theta,\phi) \in D_{in}$ \\
$V_o$ & ; $(r,\theta,\phi) \in D_{out}$
\end{tabular}
\right.
$$
and
$$
m^*(r,\theta,\phi) = \left\{
\begin{tabular}{ll}
$m^*_{in}$ & ; $(r,\theta,\phi) \in D_{in}$ \\
$m^*_{out}$ & ; $(r,\theta,\phi) \in D_{out}$
\end{tabular}
\right.
$$

The analytical solution of Eq. (\ref{eq1}) is sought in the closed form
\be
\label{serie}
\Psi = \sum_i C_i \; \Psi^{(o)}_i,
\ee
where the set of functions $\{\Psi^{(o)}_i\}$ is a complete set of functions in the 3D domain $D$
determined by the semi-sphere and its explicit representation can be found in Ref.
\onlinecite{jpacm}. With the former expansion the functions $\Psi$ satisfy the boundary condition
of infinite barrier in the contour $C_1$ because the set of functions $\{\Psi^{(o)}\}$ does. On the
other hand, Eq. (\ref{eq1}) and the corresponding solution given by Eq. (\ref{serie}) are written
for the whole domain $D$. It guarantees that the matching conditions at the contour $C_2$ are also
satisfied, but only at those points where it is well-defined the derivative of the wavefunction.
This does not occur at the corners of the contour $C_2$ where the differentiation with respect to
the normal to the surface presents a discontinuity and, generally speaking, the problem is then not
well-defined \cite{courant}. The obtained eigenvalues constitute only an estimation of the solution
for the real problem but we accept this solution as a better one than the values reported by the
infinite barrier case. The present formalism has been applied in previous works \cite{goff}, but
not explicit analysis has been done about the method used and the fulfillment of the matching
conditions between the internal and the external domain.

\begin{figure} 
\vspace*{-1cm}
\centerline{\psfig{file=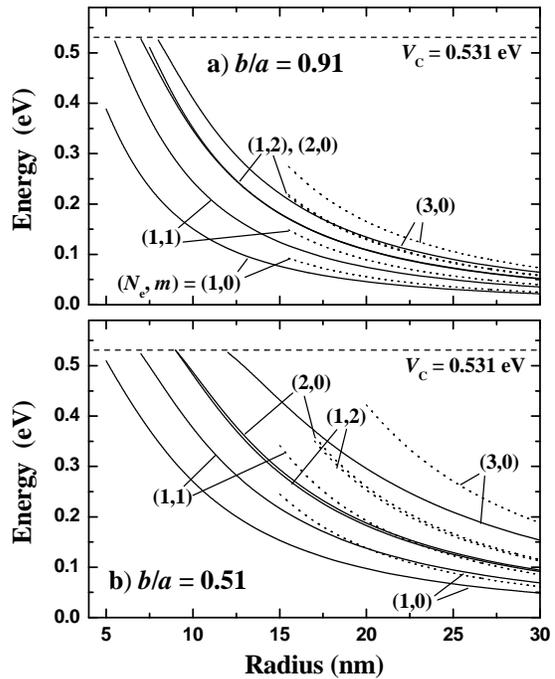,width=80mm}}
\caption{First five electronic levels in the conduction band for an InAs/GaAs quantum lens as
a function of the lens radius. a) $b/a=0.91$, b) $b/a=0.51$. The calculation is done taken
$\Delta/R=0.3$ (solid lines) for the finite barrier model. Dotted lines are the results considering
the infinite potential barrier model \cite{jpacm}. The conduction confined potential $V_c$ is
represented by dashed lines.}
\label{fig:Edim_radio}
\end{figure}

In order to visualize the results of the above described model, two different quantum lens
geometries of InAs/GaAs have been considered. In Fig. \ref{fig:Edim_radio} are shown the first five
electronic levels as a function of the lens radius. Panel a) is for $b/a=0.91$ while panel b) is
for $b/a=0.51$. The levels calculated within the hard wall model \cite{jpacm} are shown by dotted
lines while the results obtained using the finite barrier model are described by solid lines. For
the finite barrier case, it has been used a $500 \times 500$ matrix and a value of $\Delta/R=0.3$
to guarantees that the absolute and relative errors introduced by the contour  $C_1$ and the
cut-off of Eq. (\ref{serie}) in the values obtained for the energy levels are less than 10$^{-3}$.
According to the results of the Fig. \ref{fig:Edim_radio}, it can be seen that the infinite barrier
is a good approximation for radius of the order of 20 nm or higher when $b/a=0.91$. As $b/a$
decreases, the values of the energy levels increase and the effects of the finite barrier become
important. Then, it is necessary to consider the finite barrier effects to get better accuracy for
the energy levels distribution at the same range of radius.

\section{Results and discussion}
\label{effect-of-the-pressure}

\begin{figure} 
\vspace*{-1cm}
\centerline{\psfig{file=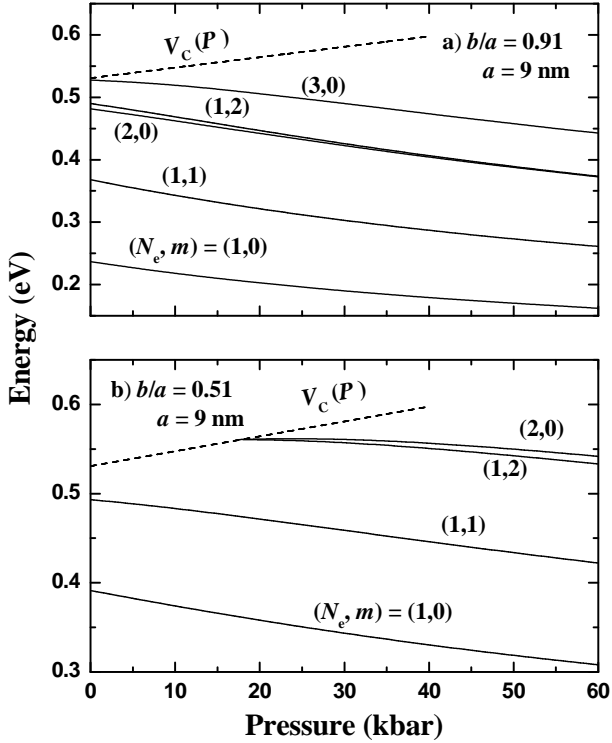,width=90mm}}
\vspace*{-0.5cm}
\caption{First energy levels in the conduction band as a function of the pressure.
a) $b/a=0.91$ and $a=9$ nm. b) $b/a=0.51$ and $a=9$ nm. In both panels the confined potential
energy it is shown by dashed lines.}
\label{fig:EdimCond_Pres}
\end{figure}

\begin{table}
\caption{Values of the parameters used in the calculation of the energy levels
for the InAs/GaAs quantum lens under pressure. The symbol $^*$ means that the tension effect of the
InAs embedded in the GaAs matrix was taken into account.}
\label{tabla1}
\begin{tabular}{lrr} \hline \hline
 & \hspace*{5mm} InAs & \hspace*{5mm} GaAs \\
\hline
$E_g^{(o)}$ (eV) & $0.533^{a,*}$  & $1.519^a$ \\
$\alpha$ (meV/kbar) & $7.7^{b,*}$ & $10.8^c$ \\
$\kappa$ ($\times 10^{-4}$ eV/K)  & $2.76^a$ & $5.405^a$ \\
$c$ (K) & 83$^a$ & 204$^a$ \\
$\gamma_1$ & $20.4^{d}$ & $6.85^{e}$ \\
$\gamma_2$ & $8.37^{d}$ & $2.1^{e}$ \\
$\delta$ (meV) & $380^{d}$ & $341^{d}$ \\
$2 \left( P_o^2/m_o \right)$ (eV) & $19^{d}$ & $12.9^{f}$ \\
\hline \hline
\end{tabular} \\
\medskip
\begin{tabular}{lll}
$^{a}$ Ref. \onlinecite{nota1}, &  $^{b}$ Ref. \onlinecite{nota2}, & $^{c}$ Ref.
\onlinecite{duque-jpcm},
\\
$^{d}$ Ref. \onlinecite{landolt}, & $^{e}$ Ref. \onlinecite{vlr}, & $^{f}$ Ref.
\onlinecite{trallero1998}
\end{tabular}
\end{table}

Figure \ref{fig:EdimCond_Pres} shows the variation with pressure of the first energy levels in the
conduction band for two lens configurations $b/a=0.91$ [Fig. \ref{fig:EdimCond_Pres}a)] and
$b/a=0.51$ [Fig. \ref{fig:EdimCond_Pres}b)] with $a = 9$ nm. In both figures it is included the
height of the potential barrier $V_o = V_c$ as a function of the pressure by dashed lines. The band
offset of the strained InAs/GaAs quantum lens was taken, for the conduction (valence) band, as
54$\%$ (46$\%$) of the total band difference \cite{duque-jpcm}.

\begin{figure}
\vspace*{2cm}
\centerline{\psfig{file=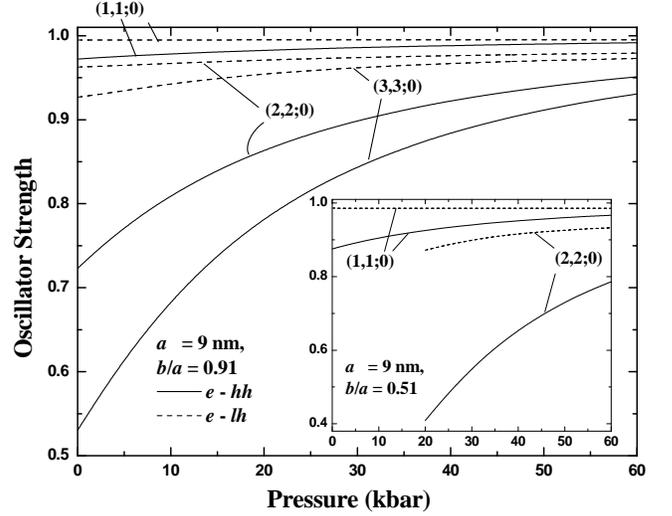,width=90mm}}
\vspace*{-2cm}
\caption{Oscillator Strength as a function of the pressure for electron-hole transitions
$(N_e,N_h;m)$ with $N_e=N_h$ and $m=0$. In solid line are shown the transitions from heavy holes
states while in dashed lines from light hole states. The configuration lens is $a=9$ nm with
$b/a=0.91$. In the inset the same radius and $b/a=0.51$.}
\label{fig:OS}
\end{figure}

It is interesting to note the decreasing of the values of the conduction energies levels with the
increasing of $P$, in spite of the increasing of the barrier height according to Eq. (\ref{Egap})
and the decreasing of the dot radius following Eq. (\ref{change_of_volume}). This effect is very
robust for both configurations shown in Fig. \ref{fig:EdimCond_Pres}. This fact is a direct
consequence of the strong influence of the variation of the electron mass with $P$ according to Eq.
(\ref{mass_electron}). The same will happen for the light hole (see Eq. (\ref{mass_ll})). In the
case of Fig. \ref{fig:EdimCond_Pres} b), due to the strong lens deformation, only two electron
states are present at $P=0$ [(1,0) and (1,1)] and two more appear at $P \approx 20$ kbar.

\begin{figure}
\centerline{\psfig{file=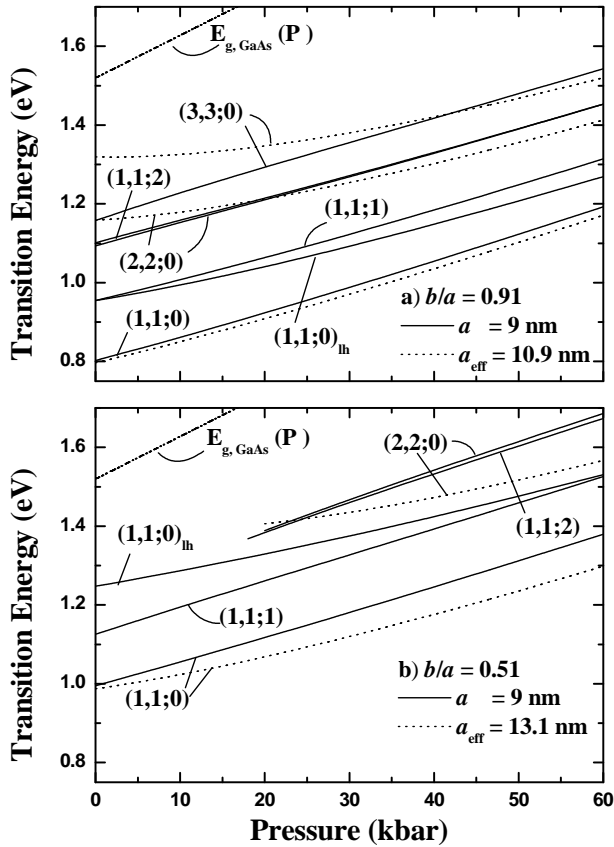,width=90mm}}
\vspace*{0.5cm}
\caption{First allowed optical transition energies of two InAs/GaAs quantum lens as a
   function of the pressure with $a = 9$ nm. a) $b/a = 0.91$. b) $b/a = 0.51$. Solid lines are obtained
   considering finite barrier while dotted lines represent the result with infinite barrier using an
   effective radius $a_{eff}$ as a fitting parameter at $P=0$.}
\label{fig:Pres}
\end{figure}

In Fig. \ref{fig:OS} it is shown the OS as a function of the  pressure, for the first electron-hole
states with $m=0$. Each line is labelled according to the corresponding transition $(N_e,N_h;m)$,
which means a transition between the valence band states $N_h$ to the conduction band states $N_e$.
Only those states with $N_e=N_h$ are shown, because they present the stronger values of the OS. The
values of the OS corresponding to the transitions with $N_h \ne N_e$ are very small and decrease
with the pressure. There are shown by solid lines the transitions from heavy hole valence band and
by dashed lines from light hole valence band, for comparison. The configuration used was $a=9$ nm
and $b/a=0.91$ while in the inset the same radius it is used, but a more confined region with
$b/a=0.51$. It can be seen that as the number $N$ increases, the OS value decreases for a given
pressure. Nevertheless, the OS for the light hole states have larger value than those corresponding
to the heavy hole contributions. From the figure follows that as the ratio $b/a$ decreases ($b/a =
0.51$ in the inset) and keeping constant the lens radius, the value of the OS is reduced. This
result is a direct consequence of the spatial confinement. Since the energy levels rise with the
confinement (see Fig. \ref{fig:EdimCond_Pres}) the wave functions are less localized and the
overlapping of the states with different effective masses decreases. Finally, it can be seen the
increasing of the OS as $P$ increases. Hence, we can conclude that at high pressure the main
contribution to the observed PL lines comes from these states with $N_e=N_h$.

\begin{figure}
\vspace*{2cm}
\centerline{\psfig{file=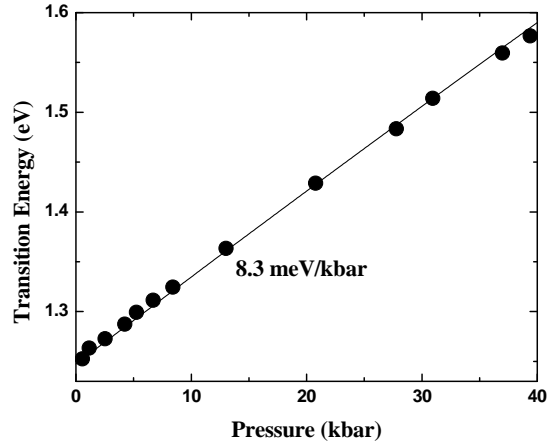,width=90mm}}
\vspace*{-2cm}
\caption{Ground state transition energy for a InAs/GaAs SAQD structure. Solid circles are data
the reported in Refs. \onlinecite{Itskevich97} and \onlinecite{Itskevich98}. Solid line represents
the theoretical calculation for a lens shape geometry with radius $a=6.4$ nm at $P=0$ and
$b/a=0.34$.}
\label{fig:Fitting}
\end{figure}

Taking in consideration the selection rules above obtained, in Fig. \ref{fig:Pres} is shown, in
solid lines, the variation with the pressure of the first allowed optical transition energies for
$N_e = N_h$ and two InAs/GaAs quantum lens configuration. In Fig. \ref{fig:Pres} a) it is
considered $b/a=0.91$ while in Fig. \ref{fig:Pres} b) we have a stronger deformed lens with
$b/a=0.51$. In both cases we used $a=9$ nm, $\Delta/R = 0.3$ and a $500 \times 500$ matrix in the
diagonalization procedure. The dashed line is the variation of the energy gap of the GaAs
surrounding medium as a function of the pressure. In general, all transitions correspond to the
electron-heavy hole states, except $(N_e,N_{lh};m)=(1,1;0)_{\mbox{lh}}$, which corresponds to the
electron-light hole state. As can be seen, the values of the transition energy for these states
always increases with pressure. This behavior is due to the InAs energy gap dependence on $P$
according to Eq. (\ref{Egap}) \cite{duque-jpcm,ahr-pssb-aceptado}. In the case of Fig.
\ref{fig:Pres} b), the stronger confinement provokes an increases of the corresponding energy
levels and at $P=0$ we are in presence of only two transitions, while transitions (1,2;2) and
(2,2:0) appear for $P > 20$ kbar.

On the other hand, each panel shows, in dotted line, the calculation using the infinite potential
barrier model. Here, we introduced an effective radius $a_{eff}$ as a fitting parameter at $P=0$
for the ground transition energy ($1,1;0$). In Fig. \ref{fig:Pres} a) the fitting procedure is
reasonable good for the electron-hole ground state (1,1;0) for all values of $P$ but it becomes
worse for the excited ones. In the case of Fig. \ref{fig:Pres} b), the implementation an $a_{eff}$
at $P=0$ does not work properly even for the (1,1;0) transition and it does not have sense for the
(2,2;0) electron-hole state transition. This result can be understood taking in consideration that
Fig. \ref{fig:Pres} b) corresponds to the strong confinement regime and the infinite barrier model
break down. It the case of a quantum lens with higher values of $b/a$ and radius (soft confinement
regime), the energy levels are deep enough in the well. Then, the finite value of the band offset,
and its dependence with the pressure, will have no appreciable influence on the transitions
energies and the model of infinite barrier will be suitable for characterize the system.

\section{Comparison with experimental results.}
\label{section-fitting}

In spite of the limitation of the present simulation we explored its viability  comparing with
available  experimental data. Figure \ref{fig:Fitting} shows, in solid circles, the experimental
transition energy for the electron-hole ground sate as a function of the pressure for an InAs/GaAs
SAQD reported in Refs. \onlinecite{Itskevich97} and \onlinecite{Itskevich98}. We have calculated
the ground state transition energy ($N_e=1,N_{hh}=1; m=0$) in terms of the applied hydrostatic
pressure. To fit the data in our calculation we fixed the ratio $b/a=0.34$ independent of $P$ and
used the value of $a=6.4$ nm at $P=0$. The theoretical result is shown in solid line. Note that the
value of $b/a$, but not the radius $a$, here employed match very well with the estimation given in
Ref. \onlinecite{Itskevich99}.

Notice from the Fig. \ref{fig:Fitting} that the value of the linear pressure coefficient is
$\mbox{d}E/\mbox{d} P = 8.3$ meV/kbar. In connection with this, the variation of the band offset
and the change of the effective masses with $P$ are the responsible of the obtained value. Also, it
is important to remark that the input used value of the linear pressure coefficient in Table
\ref{tabla1} for the InAs is linked to the good match between the simulation and the experimental
data. If the typical bulk value of 100 meV/kbar for InAs is used the agreement goes down
drastically. The consideration of the built-in strain of InAs/GaAs in the framework of
two-dimensional layer model is one of key factor that allows the correct characterization of the
fundamental optical emission as function of the hydrostatic pressure.

\section{Conclusions}
\label{conclusions}

It was developed a model which allows us to include the pressure dependence of the volume, the band
offset and the effective masses in the calculation of the energy levels at the $\Gamma$ point of
the Brillouin zone in SAQDs with lens shape geometry. It was outlined the importance of the
involved parameters on the electronic states and on optical emission for SAQDs under hydrostatic
pressure ranging between 0 and 40 kbar. Also, we have delimited the validity of the infinite
potential barrier model for the evaluation of the energy levels as a function of the hydrostatic
pressure in quantum lens.

The strong strain effect of InAs/GaAs on the gap energy and pressure coefficient can be considered
under two-dimensional layer approximation guarantying a reliable description of the fundamental PL
line as a function of the applied pressure.

The present model is a qualitative one which needs to be improved in order to include the more
complex effects present in such structures such as the real strain field in dots with lens
symmetry, the peculiarity of the valence bands in III-V semiconductors and excitonic effects.

\section*{Acknowledgement}

C. A. Duque would like to thank the Colombian COLCIENCIAS Agency and CODI-Universidad de Antioquia,
for partial financial support. This work has been partially supported by the Excellence Center for
Novel Materials - ECNM, under Colciencias contract No. 043-2005. The author is also grateful to Dr.
Anna Kurczy\'nska for useful discussion. A. H. Rodr\'{\i}guez would like to thank Dr. Rub\'en
Barrera P\'erez for fruitful comments and to the Department of Solid State, Institute of Physics,
UNAM, where part of this work was carried out. The authors also thank Dr. Nelson Porras-Montenegro
for important discussions.


\begin{thebibliography}{10}

\bibitem{Stranski-Krastanow}
I.~N. Stranski and L. Krastanow, Sitzungsber. Akad. Wiss. Wien {\bf 146},  797
  (1938).

\bibitem{hawrylak-book}
L. Jacak, P. Hawrylak, and A. Wojs, {\em Quantum dots} (Springer-Verlag,
  Berlin, 1998).

\bibitem{Grundmann-book}
D. Bimberg, M. Grundmann, and N.~N. Ledentsov, {\em The quantum dot
  heterostructures} (Wiley, Chichester, 1999).

\bibitem{cusack96}
M.~A. Cusack, P.~R. Briddon, and M. Jaros, Phys. Rev. B {\bf 54},  R2300
  (1996).

\bibitem{cusack97}
M.~A. Cusack, P.~R. Briddon, and M. Jaros, Phys. Rev. B {\bf 56},  4047
  (1997).

\bibitem{grundmann95}
M. Grundmann, O. Stier, and D. Bimberg, Phys. Rev. B {\bf 52},  11969  (1995).

\bibitem{zunger99}
Lin-Wang and A. Zunger, Phys. Rev. B {\bf 59},  15806  (1999).

\bibitem{pryor98}
C. Pryor, Phys. Rev. B {\bf 57},  7190  (1998).

\bibitem{grundmann99}
O. Stier, M. Grundmann, and D. Bimberg, Phys. Rev. B {\bf 59},  5688  (1999).

\bibitem{zunger01}
A. Zunger, Phys. Stat. Sol. (b) {\bf 224},  727  (2001).

\bibitem{mervyn}
M. Roy and P.~A. Maksym, Phys. Rev. B {\bf 68},  235308  (2003).

\bibitem{hawrylak96-1}
A. Wojs, P. Hawrylak, S. Fafard, and L. Jacak, Phys. Rev. B {\bf 54},  5604
  (1996).

\bibitem{eaglesham90}
D.~J. Eaglesham and M. Cerullo, Phys. Rev. Lett. {\bf 64},  1942  (1990).

\bibitem{forchel96}
A. Forchel {\it et~al.}, Semicond. Sci. Technol. {\bf 11},  1529  (1996).

\bibitem{hapke99}
I. Hapke-Wurst {\it et~al.}, Semicond. Sci. Technol. {\bf 14},  L41  (1999).

\bibitem{zhu98}
J.~H. Zhu, K. Brunner, and G. Abstreiter, Appl. Phys. Lett. {\bf 72},  424
  (1998).

\bibitem{liao98}
X.~Z. Liao {\it et~al.}, Phys. Rev. B {\bf 58},  R4235  (1998).

\bibitem{moison94}
J.~M. Moison {\it et~al.}, Appl. Phys. Lett. {\bf 64},  196  (1994).

\bibitem{arezky01}
A.~H. Rodr{\'{\i}}guez, C. Trallero-Giner, S.~E. Ulloa, and J.
  Mar{\'{\i}}n-Antu{\~n}a, Phys. Rev. B {\bf 63},  125319  (2001).

\bibitem{zou99}
J. Zou, X.~Z. Liao, D.~J.~H. Cockayne, and R. Leon, Phys. Rev. B {\bf 59},
  12279  (1999).

\bibitem{even}
J. Even and S. Loualiche, J. Phys. A: Math. Gen {\bf 36},  11677  (2003).

\bibitem{ngo}
C.~Y. Ngo, S.~F. Yoon, and W.~J. Fan, Phys. Rev. B {\bf 74},  245331  (2006).

\bibitem{lew}
L.~C. L.~Y. Voon and M. Willatzen, J. Phys.: Condens. Matter {\bf 14},  13667
  (2002).

\bibitem{AHR2006pss1}
A.~H. Rodr\'{\i}guez and L. Meza-Montes, Phys. Stat. Sol.(b) {\bf 243},  1276
  (2006).

\bibitem{tang}
N.~Y. Tang, X.~S. Chen, and W. Lu, Phys. Lett. A {\bf 336},  434  (2005).

\bibitem{goni1}
F.~J. Manj\'on {\it et~al.}, phys. stat. sol. (b) {\bf 235},  496  (2003).

\bibitem{goni2}
C. Kristukat {\it et~al.}, phys. stat. sol. (b) {\bf 244},  53  (2007).

\bibitem{wei}
J.-W. Luo, S.-S. Li, and J.-B. Xia, Phys. Rev. B {\bf 71},  245315  (2005).

\bibitem{Itskevich97}
I.~E. Itskevich {\it et~al.}, Appl. Phys. Lett. {\bf 70},  505  (1997).

\bibitem{Itskevich98}
I.~E. Itskevich {\it et~al.}, Phys. Rev. B {\bf 58},  R4250  (1998).

\bibitem{Itskevich99}
I.~E. Itskevich {\it et~al.}, Phys. Rev. B {\bf 60},  R2185  (1999).

\bibitem{ma2004}
B.~S. Ma {\it et~al.}, J. Appl. Phys. {\bf 95},  933  (2004).

\bibitem{mintairov}
A.~M. Mintairov {\it et~al.}, Phys. Rev. B {\bf 69},  155306  (2004).

\bibitem{luo}
J.-W. Luo, S.-S. Li, and J.-B. Xia, Phys. Rev. B {\bf 71},  245315  (2005).

\bibitem{duque-jpcm}
C.~A. Duque {\it et~al.}, J. Phys.: Condens. Matter {\bf 18},  1877  (2006).

\bibitem{frogley}
M.~D. Frogley, J.~R. Downers, and D.~J. Dunstan, Phys. Rev. B {\bf 62},  13612
  (2000).

\bibitem{trallero-unpublished}
C. Trallero-Giner, K. Kunc, and K. Syassen, Phys. Rev. B {\bf 73},  205202
  (2006).

\bibitem{PRB-DR}
A.~H. Rodr\'{\i}guez, C. Trallero-Giner, M. Mu{\~n}oz, and M.~C. Tamargo, Phys.
  Rev. B {\bf 72},  045304  (2005).

\bibitem{jpacm}
A.~H. Rodr\'{\i}guez, C.~R. Handy, and C. Trallero-Giner, J. Phys.: Condens.
  Matter {\bf 15},  8465  (2003).

\bibitem{jap}
A.~H. Rodr\'{\i}guez and C. Trallero-Giner, J. Appl. Phys. {\bf 95},  6192
  (2004).

\bibitem{courant}
R. Courant and D. Hilbert, {\em Methods of mathematical physics} (Interscience
  Publishers, New York, 1953).

\bibitem{goff}
S.~L. Goff and B. St\'eb\'e, Phys. Rev. B {\bf 47},  1383  (1993).

\bibitem{nota1}
We have evaluated the influence of the built-in strain in InAs dots on the band
  gap following the treatments of strained layers reported in Ref.
  \cite{duque-jpcm}.

\bibitem{nota2}
The band-gap linear pressure coefficients was evaluated according the nonlinear
  elasticity theory following Refs. \cite{ma2004} and \cite{duque-jpcm}. This
  treatment explain the lower value of the $\alpha$ coefficient for strained
  InAs relative to the unstrained bulk value. The model assumes a simple strain
  two-dimensional layers.

\bibitem{landolt}
{\em Landolt B{\"{o}}rnstein Tables}, Vol.~37a of {\em Landolt-B{\"o}rnstein
  New Series, Group III}, edited by {O}. {M}adelung ed. (Springer-Verlac,
  Berlin, 1982).

\bibitem{vlr}
V. L\'opez-Richard, G.~E. Marques, C. Trallero-Giner, and J. Drake, Phys. Rev.
  B {\bf 58},  16136  (1998).

\bibitem{trallero1998}
C. Trallero-Giner, A. Alexandrou, and M. Cardona, Phys. Rev. B {\bf 38},  10744
   (1998).

\bibitem{ahr-pssb-aceptado}
A.~H. Rodr\'{\i}guez {\it et~al.}, Phys. Stat. Sol.(b) {\bf 244},  48  (2007).

\end{thebibliography}
\end{document}